\documentclass[conference]{IEEEtran}
\IEEEoverridecommandlockouts
\usepackage{subfig}
\usepackage{multirow}
\usepackage{amsmath,amssymb,amsfonts}
\usepackage{graphicx}
\usepackage{xcolor}
\usepackage{url}
\usepackage{soul}
\usepackage{siunitx}
\usepackage{lipsum}

\usepackage{hyperref}
\hypersetup{
    colorlinks=true,
    linkcolor=blue,
    filecolor=magenta,      
    urlcolor=cyan,
    pdftitle={The MIDNA Skipper-CCD Readout ASIC with Improved Performance},
    }
\begin{document}

\title{A Sub-electron-noise Skipper-CCD Readout ASIC with Improved Channel-to-channel Isolation and an Integrated Cryogenic Voltage Reference}

\author{Fabricio Alcalde Bessia, Claudio Chavez, Troy England, Hongzhi Sun, Andrew Lathrop, \\
Davide Braga, Miguel Sofo-Haro, Juan Estrada, Farah Fahim

\thanks{F. Alcalde Bessia is with Instituto Balseiro and Instituto de Nanociencia y Nanotecnología INN (CNEA-CONICET), R8402AGP San Carlos de Ba\-ri\-lo\-che, Argentina (e-mail: \protect\url{falcalde@ib.edu.ar}).}
\thanks{C. Chavez, T. England, H. Sun, A Lathrop, D. Braga, J. Estrada and F. Fahim are with the Fermi National Accelerator Laboratory, Pine \& Kirk St, Batavia, IL 60510 (e-mail: \protect\url{tengland@fnal.gov}).}
\thanks{M. Sofo-Haro is with Universidad Nacional de C\'ordoba, Comisión Nacional de Energía Atómica (CNEA) and CONICET, X5000HUA C\'ordoba, Argentina (e-mail: \protect\url{miguelsofoharo@mi.unc.edu.ar}).}}

\maketitle

\begin{abstract}
    The \emph{MIDNA} application specific integrated circuits (ASICs) are a series of skipper-CCD readout chips fabricated in a 65\,nm low-power CMOS process that implement a correlated double sampling signal processing chain based on dual-slope integrators. They are capable of working from room to cryogenic temperatures, down to \SI{84}{\K}. The present iteration of the ASIC has been fabricated including several design updates and the addition of an on-chip voltage reference, resulting in improved performance. This work presents the main vulnerabilities solved, the changes carried out, and the resulting performance benefits. Measurements with a skipper-CCD and the ASIC at \SI{140}{\K} showed that the single-electron resolution can be reached by averaging the measured charge in the analog domain using the analog pile-up technique with a readout noise as low as \SI{0.11}{e^{-}_{rms}} of equivalent charge for 1200 samples. The channel-to-channel crosstalk was also characterized showing values better than \SI{-62}{\dB}.
\end{abstract}


\section{Introduction}

Skipper Charge Coupled Devices (CCDs) are silicon, pixelated, particle detectors that are currently being used for astronomy \cite{marrufo2024, villalpando2024} and for very sensitive physics experiments, including dark matter searches \cite{barak2020sensei, aguilar2019constraints, d2020violeta, Estrada2022oscura}. Thanks to their non-destructive readout capability and high charge-transfer efficiency, the readout noise of Skipper-CCDs can be lowered by averaging multiple samples of the same charge. In \cite{tiffenberg_2017}, an impressive \SI{0.068}{e^{-}_{rms}} readout noise floor was demonstrated, allowing for precise electron counting up to thousands of electrons in each of the million pixels across the sensor. This property, in addition to the low background rate of Skipper-CCDs, makes them one of the most promising detector technology for the search for dark matter in the sub-GeV energy range. Increasing the amount of fiducial mass for the faint dark matter interactions is key, so scientific collaborations are aiming to make experiments with large volumes of Skipper-CCDs. For example, the SENSEI experiment \cite{barak2020sensei} intends to make a \SI{\approx 100}{\g} Skipper-CCD array, and the DAMIC-M experiment \cite{aguilar2019constraints} is aiming to manufacture a \SI{1}{\kg} array. The OSCURA experiment \cite{Estrada2022oscura} is aiming to assemble a \SI{10}{\kg} array, i.e. approximately 24000 Skipper-CCDs working simultaneously in a low-temperature vessel. 

Integrating 24000 Skipper-CCDs presents challenges. The single electron counting capability of Skipper-CCDs has been demonstrated with the Monsoon acquisition system \cite{tiffenberg_2017}, and then specific electronics systems were designed, like the LTA (Low-Threshold Acquisition) board \cite{CanceloLTA2021} and the analog front-end electronics developed for OSCURA \cite{chavez2022multiplexed}. However, such discrete systems cannot be scaled to thousands of channels because of the power constraints of cryogenic cooling systems and the cabling required to connect each sensor.

The MIDNA ASICs were designed to overcome the scalability issue. The current work implements four skipper-CCD readout channels in just \SI{2}{\mm\squared} at a fraction of the power and volume required by discrete electronics systems. It is also capable of working from room temperature down to \SI{84}{\K} and, therefore, it can be placed close to the sensor for in-situ signal processing and amplification. The ASIC was fabricated in a low-power \SI{65}{\nm} CMOS technology. The previous iteration \cite{midna1paper} demonstrated its low-noise capability by enabling single electron resolution. The work presented here starts with an overview of the working principles of the chip. It introduces the design weaknesses that have been addressed. Then, the design changes and improvements implemented in this version are explained in detail. Finally, an experimental demonstration of the ASIC's capabilities is shown and the performance parameters related to the design changes are compared to previous results. 

\section{ASIC Design and Improvements}

\begin{figure}
    \centering
    \includegraphics[width=\linewidth]{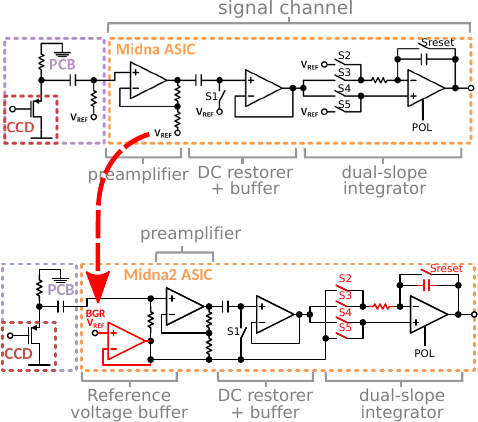}
    \caption{Comparison between the signal paths in the previous iteration (top) and the current work (bottom). Each chip includes four channels.}
    \label{fig:MIDNA1MIDNA2ChannelComparison}
\end{figure}

\begin{figure}
    \centering
    \includegraphics[width=\linewidth]{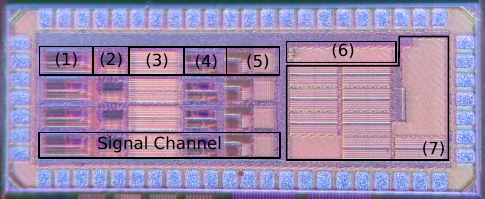}
    \caption{Photo of the present ASIC. Its size is \SI{2.5}{\mm} by \SI{1}{\mm}. The main blocks are: (1) Preamplifier, (2) reference-voltage buffer, (3) DC restorer, (4) signal buffer, (5) integrator, (6) bias generators, and (7) bandgap reference.}
    \label{fig:MIDNA2photo}
\end{figure}

The main purpose of the ASIC is to perform the correlated double sampling (CDS) of the CCD signal in the analog domain with a dual-slope integrator (DSI) circuit \cite{janesick2001scientific}. Fig. \ref{fig:MIDNA1MIDNA2ChannelComparison} shows a comparison between the signal paths of the first and the second versions of the ASIC, highlighting the changes and additions in red. 
One channel is composed of a preamplifier stage, a DC restore stage, and a reconfigurable integrator stage. A reference voltage is needed to implement a mid-rail signal ground, allowing for positive and negative excursions with a single \SI{2.5}{\V} power supply configuration. The reference voltage was set to \SI{1.1}{\V} by design, thus leaving more room for positive swings. It was designed and fabricated in a low-power \SI{65}{\nm} CMOS fabrication process using \SI{2.5}{\V} devices for voltage headroom. Design for cryogenic temperatures was assisted by the availability of cryogenic SPICE models \cite{cryomodels1,cryomodels2}. A photo of the ASIC indicating the main components is show in figure \ref{fig:MIDNA2photo}.

The chip has five real-time control inputs: DCR, INTN, INTP, RESET and POL. DCR controls the state of the switch S1 of Fig. \ref{fig:MIDNA1MIDNA2ChannelComparison}, closing it when in high state. INTN and INTP control switches S2 to S5, which set the integrator stage as an inverting and non-inverting integrator. The RESET input controls the Sreset switch, which discharges the integrator capacitor. Lastly, the POL input changes the integrator's operational amplifier offset direction, making it possible to chop the amplifier. 

\begin{figure}
    \centering
    \includegraphics[width=\linewidth]{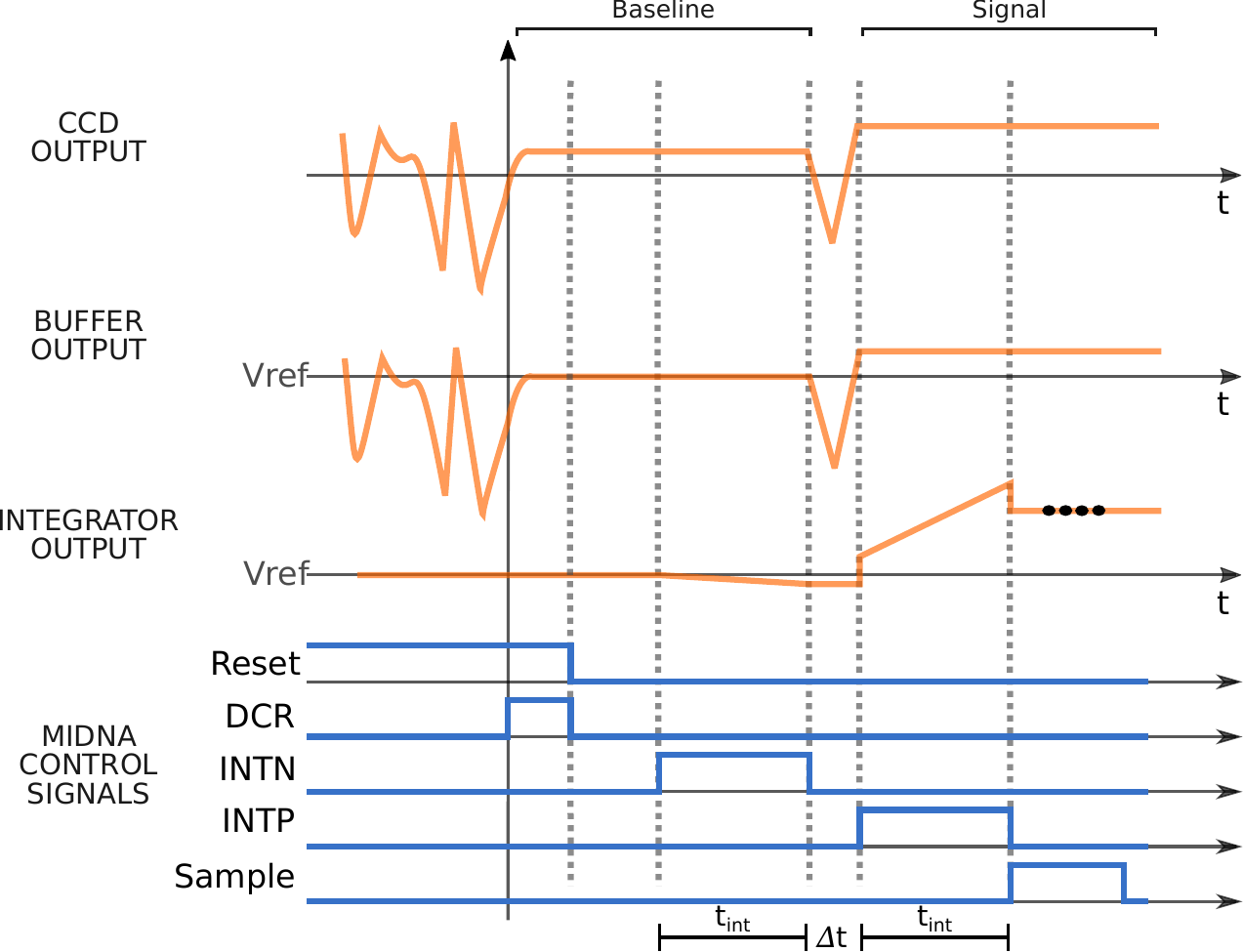}
    \caption{Timing diagram showing the control signals along with the CCD output and intermediate signals of the channel. Adapted from \cite{alcalde2023noise}.}
    \label{fig:readouttiming}
\end{figure}

Fig. \ref{fig:readouttiming} shows a timing diagram of a readout sequence. The cycle begins with the output of the CCD at the baseline level and the integrator in a reset state with the capacitor discharged. A short pulse on the DCR input forces the DC voltage at the integrator input to the reference voltage to avoid saturating the integrator output during the first integration phase if the baseline voltage is not settled to the reference voltage.
% Thanks to this short DC restore phase, the following integration of the baseline value with the integrator in a inverting configuration leads to an output close to the reference, avoiding the saturation that would happen when, due to the CCD clocking and AC coupling, the baseline move away from the reference.
After the inverting integration phase, the integration of the CCD signal level follows in the non-inverting configuration and, finally, the resulting voltage gets sampled by the acquisition system. 

Assuming that the CCD output is constant during the baseline and signal phases, at the end of the sequence the sampled voltage is \cite{alcalde2023noise}
\begin{equation}\label{eqn:Vsample}
    V_{sample}=V_{ref}+G_{pre} \frac{t_{int}}{R C} \left(V_{signal}-V_{baseline}\right),
\end{equation}
where \(V_{ref}\) is the reference voltage, \(G_{pre}\) is the preamplifier gain, \(t_{int}\) is the integration time, and \(R\) and \(C\) are the integrator resistor and capacitor values, respectively. 

When performing multiple non-destructive reads using a Skipper-CCD, the sequence of Fig. \ref{fig:readouttiming} is repeated tens or hundreds of times, moving the charge back and forth under the CCD output stage in order to obtain equivalent samples of the same charge \cite{tiffenberg_2017}. During this sequence, the readout noise in uncorrelated, but the signal is correlated. The multiple samples are then either, averaged in the digital domain \cite{CanceloLTA2021} or summed in the analog domain internally on the chip using the analog pile-up technique (\cite{midna1paper, sofoharo2021anal}), effectively reducing the input-referred readout noise by a factor of \(1/\sqrt{N}\), with \(N\) being the number of samples. In either approach to readout, the sampled voltage clearly depends on the channel \(V_{ref}\), and this creates a vulnerability because any change to this voltage directly affects the measured sample as shown in equation \ref{eqn:Vsample}. 

\begin{figure}
    \centering
    \includegraphics[width=\linewidth]{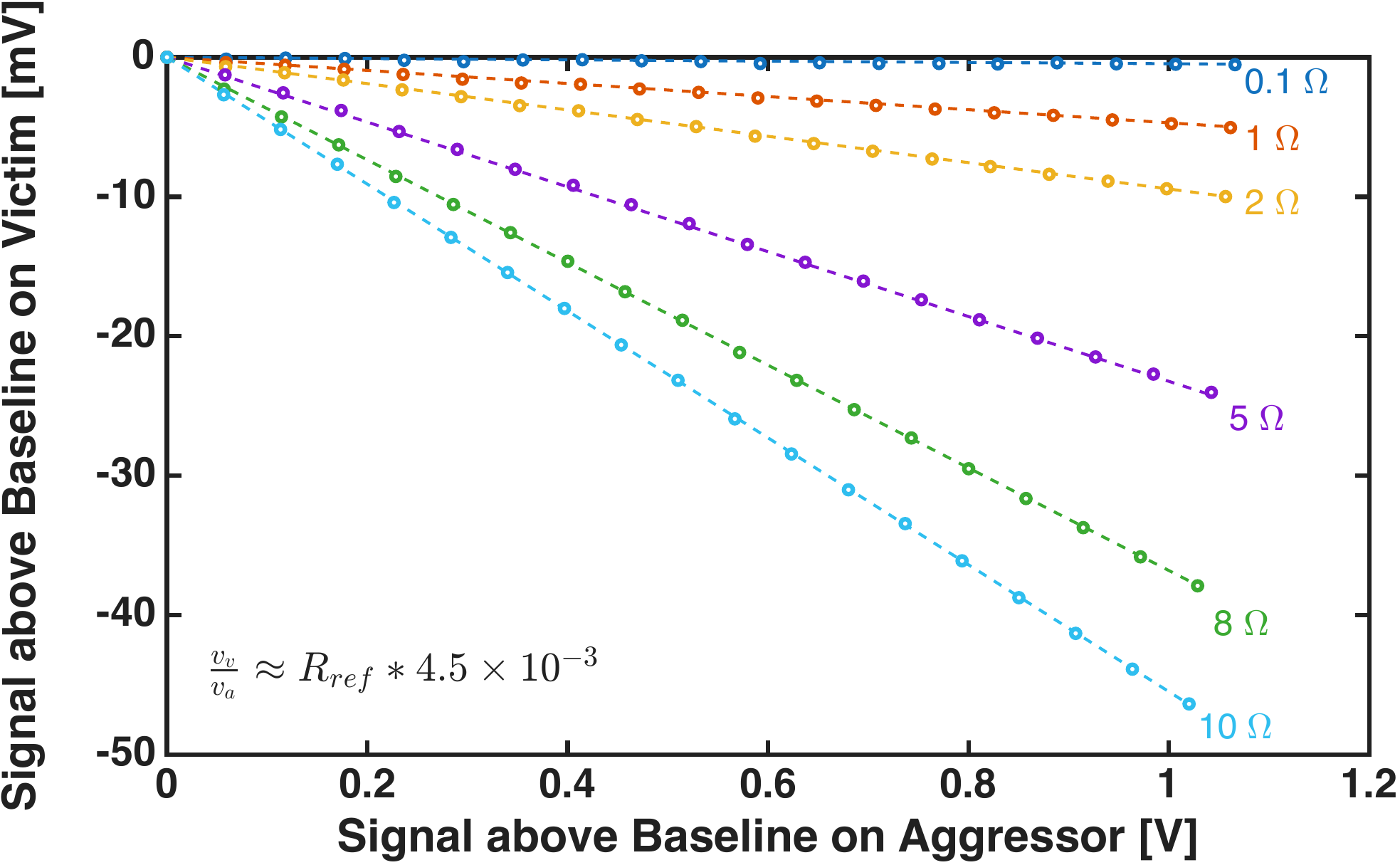}
    \caption{Simulated crosstalk between two channels of the previous chip due to shared resistance to the reference voltage.}
    \label{fig:simxtalk}
\end{figure}

In the previous version of the chip the reference voltage was supplied outside the ASIC with a commercial low-noise voltage regulator and brought on the chip through a pair of analog I/O pads. Channels zero and one shared one reference input pad, and channels two and three shared the other. The reference voltage was distributed to each one of the stages inside the channels with the lowest parasitic resistance practical. Although a large effort was put into reducing the resistance of this connection, the current flowing to the reference from each channel, mainly from the feedback resistors of the preamplifier, produced a non-negligible voltage change. Furthermore, a large signal on one channel produced a voltage change affecting the other channel on the same reference input pad as shown in Fig. \ref{fig:simxtalk}. This led to observable crosstalk in the acquired images. Since all channels are read out simultaneously, the crosstalk appears as \emph{ghost} traces in the same pixels positions as the large signal but on different channels, as will be shown on section \ref{sec:crosstalkmeasurements}. The use of a Skipper-CCD and averaging a large number of samples did not minimize the problem because of the synchronous and systematic nature of the error. 

To minimize this crosstalk, a reference voltage buffer was added on every channel of this iteration, as shown in figure \ref{fig:MIDNA1MIDNA2ChannelComparison}. This buffer was designed to isolate each channel from others signal activity. On the new ASIC the return path of the preamplifier feedback resistors current is provided by the new buffers. If there is a voltage drop on the reference path, it is isolated to just the channel itself, and it does not affect the others. The drawback is that the addition of this new block increases the power consumption per channel to \SI{6.5}{\mW}, whereas the previous iteration required \SI{4.2}{\mW}. 

% \bigskip
%\todo{INTEGRATOR OFFSET: Explain why we increased the capacitance of the integrator and added dummy switches. It is because we wanted to reduce the integration offset when reading a large number of skipper samples. There are two offset issues: 1- Main opamp offset, which is canceled out by toggling the polarity input for every skip. This also reduces the 1/f corner frequency. 2- Charge injection added each time we change the configuration from inverting to non-inverting.}

Another issue encountered in the previous iteration is the saturation of the output when adding a large number of samples using the analog pile-up technique. With this technique the integrator capacitor is discharged only once at the beginning of a pixel readout sequence and then each \emph{skip} cycle adds the resulting charge after CDS, effectively performing an addition of samples in the analog domain. At the end, the output is digitized. This technique is key for the OSCURA experiment, because it relaxes the speed constraint of the analog to digital converters \cite{Estrada2022oscura, chavez2022multiplexed}.

In the ideal case with a zero signal input, the output at the end of a CDS cycle is zero. However, offsets from the CCD and the on-chip integrator cause the output of the channel to step towards the limits of its voltage range with each CDS cycle. In general, a uniform constant added to every pixel on a CCD image is not a problem because it can be subtracted on a postprocessing step. However, when using the analog pile-up technique, this offset reduces the room for signal excursion and, therefore, induces a limit to the maximum number of samples that can be summed. A restriction to the number of summed samples also means that the input-referred noise cannot be made as low as desired.

The first method of reducing the offset involves the use of the POL input. The POL input controls switches that flip the amplifier offset polarity and can be used to cancel out the offset. Instead of switching POL between positive and negative integrations, which would inject charge affecting the CDS result, the toggling is performed between complete CDS cycles, when the integrator is in reset state. This means that always an even number of skipper samples must be taken. Thus, the residue after one CDS cycle gets canceled out on the next cycle with an inverted offset polarity. However, in the previous iteration the offset was not completely removed, leading to the signal room limitation described in the previous paragraph. After a careful analysis, two main changes were made to the integrator stage in the present work: (1) Complementary switches were added in parallel with switches S2 to S5 and Sreset; and (2) The feedback capacitance was doubled, from \SI{20}{\pF} to \SI{40}{\pF}, and the resistance lowered from \SI{75}{\kohm} to \SI{37.5}{\kohm}, keeping the same RC constant. The complementary switches were added to provide some canceling of the non-negligible charge being injected into the integrator. The larger capacitor lowered the resulting voltage from the injected charge not mitigated by the complementary switches. Finally, the layout was modified in order to improve offset symmetry. 

The last addition to the present ASIC is an on-chip low-noise bandgap voltage reference. This internal reference generates the \SI{1.1}{\V} \(V_{ref}\) voltage needed for channel operation. For an experiment with a large number of Skipper-CCDs like OSCURA, having a commercial voltage reference circuit inside the cryogenic chamber, close to the sensors, represented a challenge due to availability of cryogenic COTS, noise issues, and radiopurity. The design and implementation of the on-chip band-gap reference (BGR) is presented in section \ref{sec:bgrdesign}. 

Finally, the first iteration of the ASIC required off-chip \SI{100}{\kohm} biasing resistors connected to each one of its inputs, as shown in Fig. \ref{fig:MIDNA1MIDNA2ChannelComparison}. As with the voltage reference and towards a greater integration, these resistors were moved inside the ASIC on this new version.

\subsection{Reference Buffer Design}

The purpose of the reference voltage buffer is to isolate the global BGR distribution from the in-channel reference voltage, so that the current flowing in one channel reference does not affect the others. The reference buffer is implemented as a high-gain amplifier in a unity feedback configuration. The buffer was designed to the following specifications: (1) The output noise must be less than or equal to the input referred noise of the preamplifier to add only negligible noise to the signal path. (2) It must present a very low output impedance, so the reference voltage changes by less than the equivalent of an electron (\SI{\approx 2}{\uV}) when the maximum current is flowing to the reference buffer, i.e. the maximum signal at the preamplifier output. (3) The 3-sigma offset, including process and mismatch variations, should be less than \SI{1}{\mV}.

\begin{figure}
    \centering
    \includegraphics[width=\linewidth]{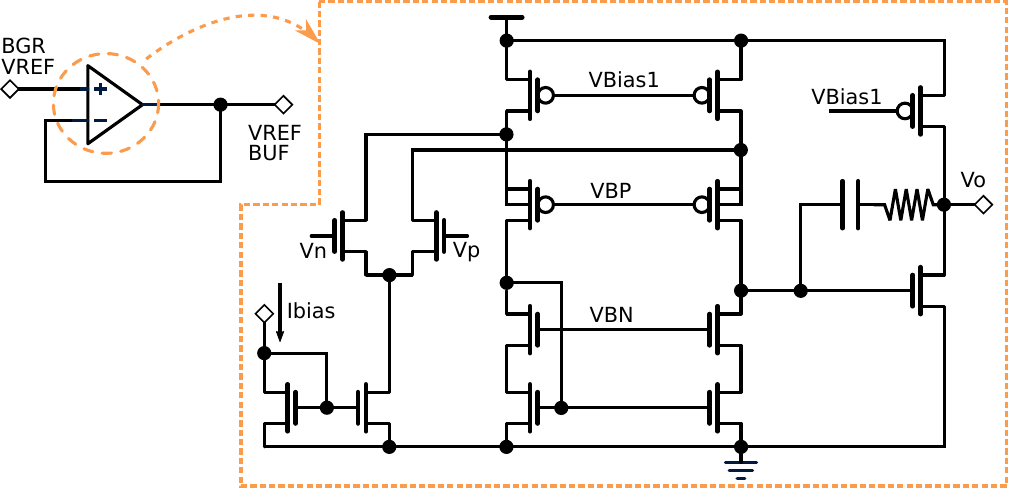}
    \caption{Schematic of the reference buffer implemented for each channel.}
    \label{fig:vrefbuffer}
\end{figure}
With these design specifications in mind, the buffer of Fig. \ref{fig:vrefbuffer} was implemented. It uses MOS gates as inputs for a very high DC input impedance, so it does not pull current from the bandgap reference. The simulated output resistance is \SI{6}{\m\ohm} and the 3-sigma offset is \SI{0.7}{\mV}, with a power consumption of \SI{1.9}{\mW}.

On a physical level the buffer was laid out close to the preamplifier feedback loop. Separate metal paths were used to connect the buffer to the preamplifier and the integrator to reduce the effect of the current flowing through the return path of the preamplifier feedback loop on the integrator reference voltage. 

\subsection{Bandgap Reference Design}\label{sec:bgrdesign}
The BGR is a standard low-voltage bandgap reference \cite{razavi2021design}, featuring an output divider and low-pass filter for noise suppression, optimized for cryogenic operation. A similar architecture, based on lateral pnp devices, has been validated at \SI{77}{\K} for the Deep Underground Neutrino Experiment (DUNE) \cite{grace2022ColdADC}, however a completely new design was required for this work to satisfy the target noise and output-voltage specifications: (1) a nominal DC output voltage of \SI{1.1}{\V} with a maximum 10\% deviation across the whole temperature range (\SI{77}{\K} to \SI{300}{\K}) and (2) less than \SI{3}{\nV\per\sqrt{\Hz}} white noise and less than \SI{500}{\nV\per\sqrt{\Hz}} at \SI{1}{\Hz} noise, identical to the reference buffer.  Monte Carlo simulations showed a standard deviation of the output voltage due to process and mismatch variations equal to \SI{1.6}{\mV} with a mean equal to \SI{1.1}{\V} and a maximum power consumption of \SI{450}{\uW}. The BGR occupies \SI{0.45}{\mm\squared}, of which only 20\% corresponds to active circuitry. The fixed overall chip area allowed the remaining space to be utilized for large MIM and MOM capacitors in the RC filter, enhancing noise suppression.

\section{Measurements}

%\todo{Test setup, Skipper-CCDs, measurement results (BGR output vs. temperature, BGR noise, integrator offset, image acquisition, crosstalk).}

This section presents the measured performance of this work. When appropriate, performance comparisons between this work and the previous iteration will be presented. The experiments were performed using a Skipper-CCD designed by MicroSystem Labs of Lawrence Berkeley National Laboratory (LBNL) \cite{tiffenberg_2017,holland2003fully,CERVANTESVERGARA2023167681}. It is a p-channel CCD fabricated on high-resistivity float-zone refined n-type silicon. The substrate is \SI{675}{\um} thick with a resistivity of approximately \SI{10}{\kohm\cm}. It reaches full depletion with a bias voltage of \SI{70}{\V}. The sensor is divided in \(1278\times1058\) square pixels with a pixel size of \SI{15}{\um}. The sensitivity of this skipper-CCD is \SI{1.8}{\uV\per{}e^-} in the setup conditions of the present work. The Skipper-CCD and the ASIC were inside a vacuum chamber at \SI{140}{\K}, whereas the auxiliary electronics and acquisition board were outside, at room temperature. 
The ASIC gain was set to its minimum (preamplifier gain: \SI[per-mode = symbol]{10}{\V\per\V}; Integration constant: \SI{0.5}{\V\per\us}) to avoid saturation, unless otherwise noted. The CCD output was biased with a \SI{100}{\kohm} resistor and AC coupled to the ASIC with a \SI{100}{\pico\farad} capacitor. 
The Low-Threshold Acquisition (LTA) board \cite{CanceloLTA2021} was used as the main controller of the system, providing clock generation according to a software defined sequence and the digitization of the ASIC outputs, for which an 18-bit \SI{15}{MSPS} ADC on the LTA board is used. 

The results presented in this section were obtained with an analog pile-up readout sequence, using an integration time $t_{int}$ of \SI{13.3}{\us} with a skipper period of \SI{52}{\us}. The pixel readout speed was limited by the sensor and cabling.

The single electron resolution capability of this work was evaluated by acquiring several images from a Skipper-CCD and increasing the number of averaged samples in each one up to 1200. Before starting each acquisition, the accumulated charge on the CCD array was zeroed by using a quick cleaning sequence. The size of stored images was 50 by 1200 pixels, meaning that an active CCD region of 50 rows by 1058 columns was read with the addition of an overscan region of 142 columns. This overscan region is useful to evaluate the readout noise of the system independently from any charge generated in the CCD active region. Although the charge in the overscan should be zero, some thermally generated charges can be accumulated during the readout time \cite{sensei_2022}. For this reason, when reading the overscan region the charges were moved in opposite direction, away from the the readout node.

\begin{figure}
    \centering
    \subfloat[Pixel histograms showing single electron resolution.]{
        \includegraphics[width=\linewidth]{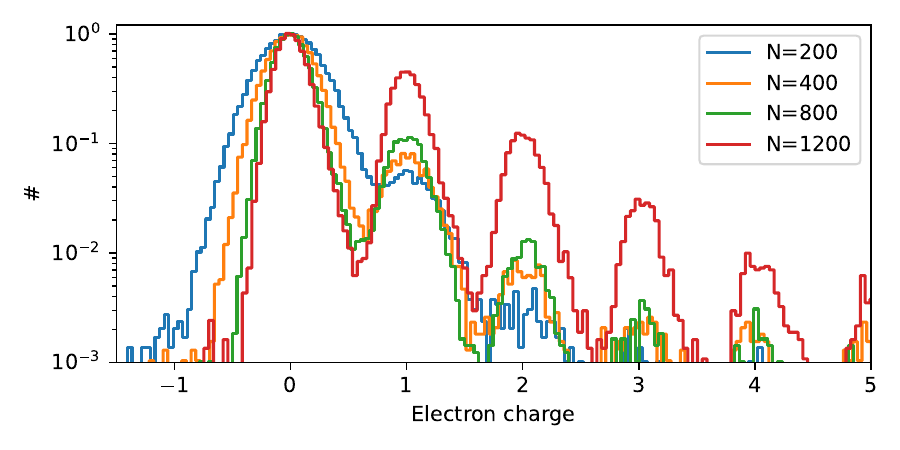}
        \label{fig:singleelectronhistograms}
    }

    \subfloat[Readout noise reduction. The dashed line shows the reduction trend with a factor $1/\sqrt{N}$.]{
        \includegraphics[width=\linewidth]{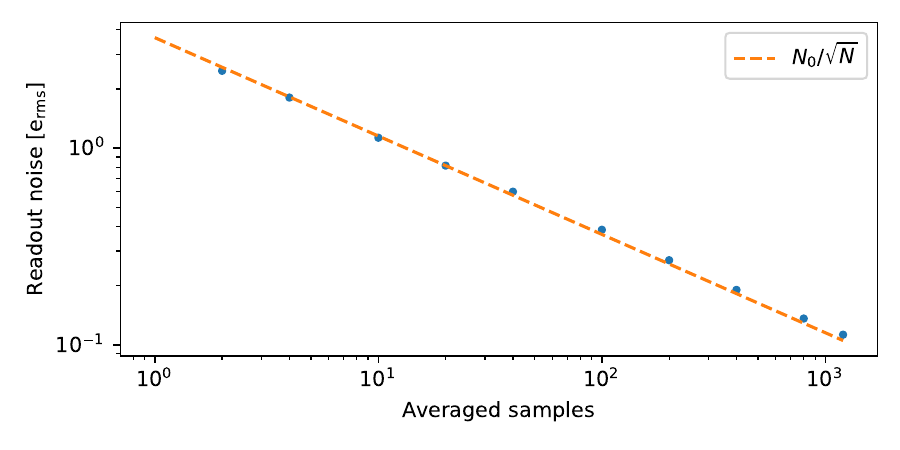}
        \label{fig:readoutnoisevsnsamp}
    }
    \caption{Single electron resolution using a Skipper-CCD and the ASIC in the analog pile-up readout mode.}
    \label{fig:singleelectrondemonstration}
\end{figure}

The pixel histograms corresponding to the active CCD region are shown in Fig. \ref{fig:singleelectronhistograms}. In order to translate the x-axis from digital units to electrons charge, a multi-Gaussian function of the form 
\[f(x, N) = \sum_{i=0}^{N-1} A_i \times \mathrm{exp}\left\{-\frac{(x-G\times{}i-O)^2}{2 \sigma^2}\right\}\]
was fit to the histogram, obtaining the gain \(G\), the standard deviation \(\sigma\), the offset \(O\), and the \(N\) multipliers \(A_i\). Then, the gain parameter, which is the peak-to-peak distance was used to rescale the x-axis by 138.45\,$\mathrm{ADU/e^{-}}$ per number of averaged samples. To ease with comparison, the histogram height was normalized by dividing by the multiplier \(A_0\).

As seen in Fig. \ref{fig:singleelectronhistograms}, increasing the number of averaged samples reduces the readout noise, which is shown by the narrowing width of each Gaussian peak. The peaks for single electron counting are clearly observable from the results, proving the acquisition with current chip, including its analog pileup feature for summing in the analog domain, enable single electron resolution and charge counting. 

The readout noise versus the number of averaged samples is shown in Fig. \ref{fig:readoutnoisevsnsamp}. The readout noise was obtained as the standard deviation of a Gaussian fit of the histograms calculated on the overscan region, where no charge is present. Then, the extracted standard deviations were divided by the system gain, obtaining the readout noise in electrons RMS. As shown in Fig. \ref{fig:readoutnoisevsnsamp}, increasing the number of averaged samples reduces the readout noise a factor \(1/\sqrt{N}\), verifying the theoretical prediction. The lowest equivalent noise charge presented here is \SI{0.11}{e^{-}_{rms}} when summing 1200 samples in the analog domain with the analog pile-up technique. %\todo{Compare the readout noise with and without CCD. FAB: I COULDN'T MAKE IT WORK}.

\subsection{Crosstalk measurements}\label{sec:crosstalkmeasurements}

\begin{figure}
    \centering
    \subfloat[Region of interest]{
        \includegraphics[width=0.97\linewidth]{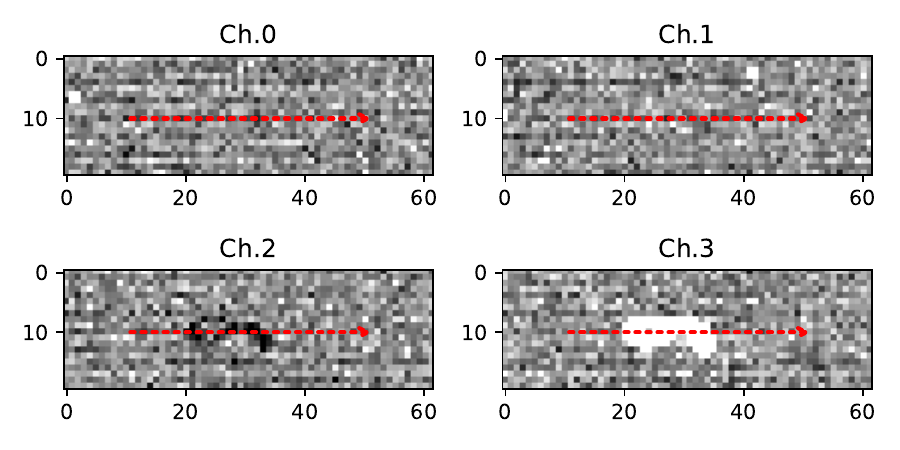}
        \label{fig:crosstalkMIDNA1_images}
    }

    \subfloat[Pixel data across the dotted line for each channel.]{
        \includegraphics[width=0.97\linewidth]{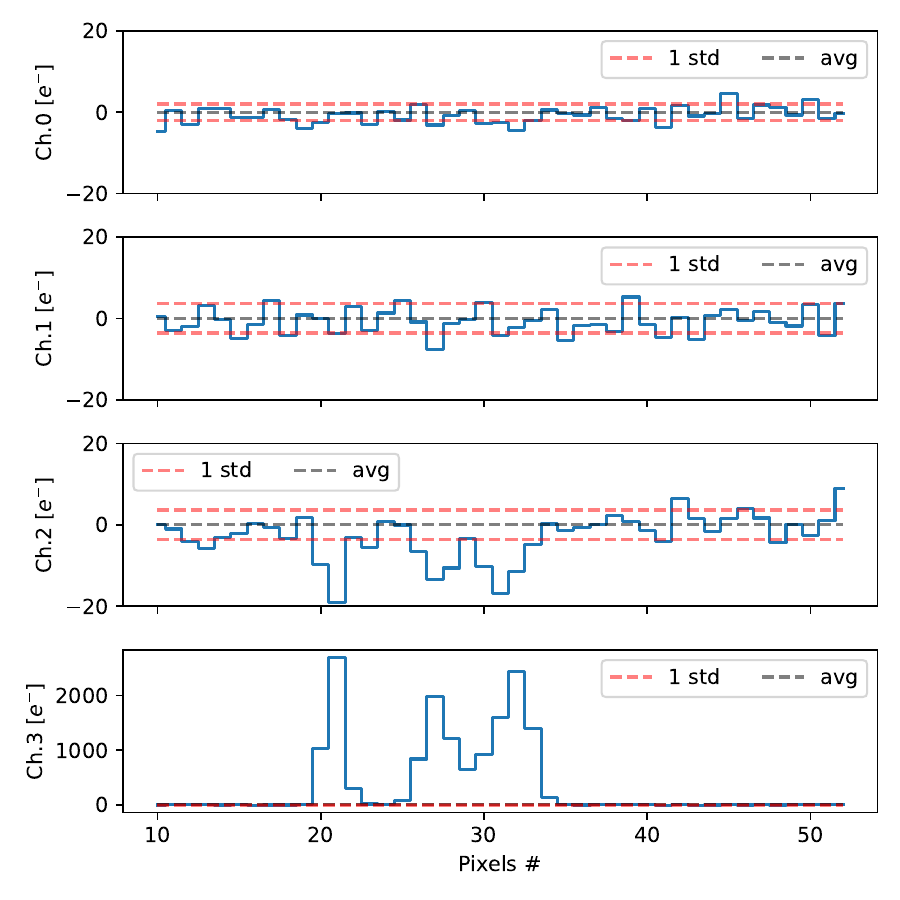}
        \label{fig:crosstalkMIDNA1_pixelvalues}
    }

    \subfloat[Victims pixel value vs. large signals on aggressor channels.]{
        \includegraphics[width=0.97\linewidth]{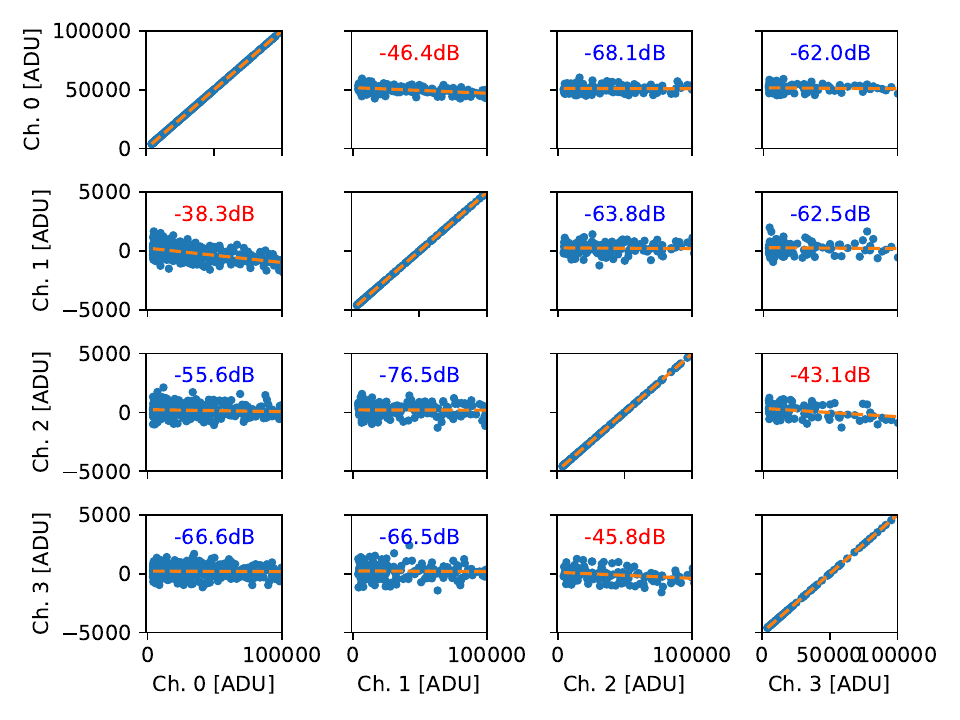}%
        \label{fig:crosstalkMIDNA1_correlation}
    }
    
    \caption{Crosstalk evaluation of the previous iteration of the ASIC. A large signal on channel 3 due to a particle interaction acts as an aggressor and channel 2 is the victim.}
    \label{fig:crosstalkMIDNA1}
\end{figure}

The crosstalk between channels was evaluated first by using ionizing events, i.e. cosmic rays, detected by the CCD during operation. Since all channels are read concurrently, a large signal in one channel could be coupled to others, affecting their final result. For the previous iteration the crosstalk was specially noticeable between channels sharing the same reference input pad, i.e. between channels 0-1 and 2-3. Fig. \ref{fig:crosstalkMIDNA1_images} serves as a good example as it shows a particle trace on channel 3 appearing as a \emph{ghost} on channel 2. The amount of charge on each pixel over the dashed red lines are plotted in figure \ref{fig:crosstalkMIDNA1_pixelvalues} and helps the quantification of the crosstalk, which can be measured as \(20\times{}log(|A_2|/|A_3|)\), with \(A_x\) the amplitude of channel \(x\). In this case, the crosstalk is equal to \SI{-42}{\dB} and corresponds to approximately \SI{2}{\ohm} of shared resistance to the reference voltage as shown in Fig. \ref{fig:simxtalk}, which matches the expected resistance from the shared metal routes.

Crosstalk was also measured by making scatter plots where pixel values of the victim channels are on the y-axis and large signals of the aggressor channels are on the x-axis, for all combinations of channels as shown in Fig. \ref{fig:crosstalkMIDNA1_correlation}. There is a negative slope with a linear trend for channels 0 and 1, indicating a non-negligible correlation coefficient. The same happens for channels 2 and 3, whereas in the other combinations the correlation is significantly closer to 0. A linear regression was performed over the data, obtaining the relation between the victim and the aggressor signal amplitude given by the slope of linear fit. The crosstalk parameters, in \si{\dB}, calculated from the slopes are shown in the same Fig. \ref{fig:crosstalkMIDNA1_correlation}.

\begin{figure}
    \centering
    \subfloat[Region of interest.]{
        \includegraphics[width=0.97\linewidth]{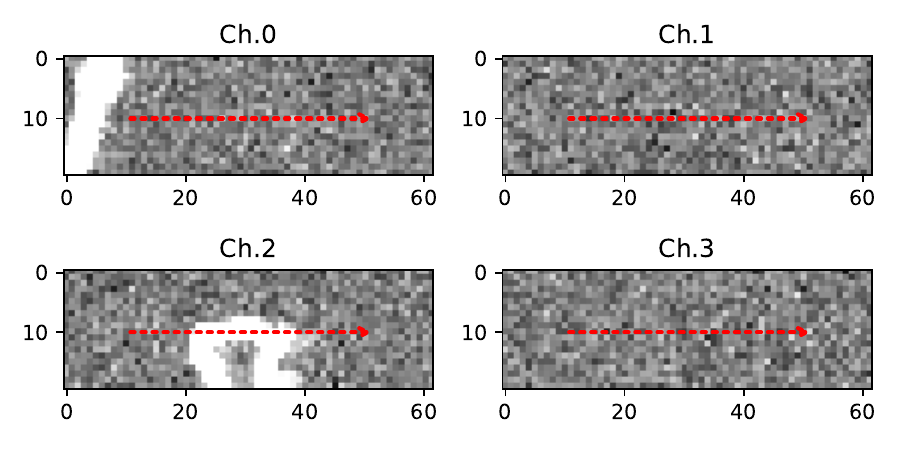}
        \label{fig:crosstalkMIDNA2_images}
    }

    \subfloat[Pixel data across the dotted line for each channel.]{
        \includegraphics[width=0.97\linewidth]{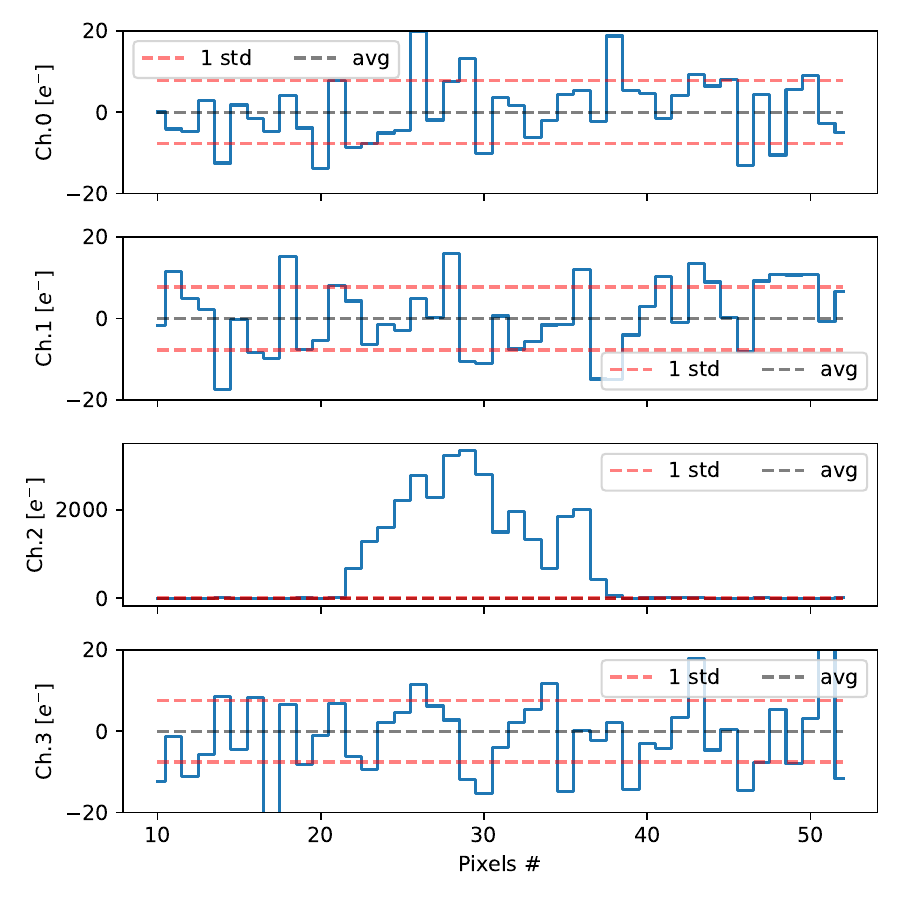}
        \label{fig:crosstalkMIDNA2_pixelvalues}
    }

    \subfloat[Victims pixel value vs. large signals on aggressor channels.]{
        \includegraphics[width=0.97\linewidth]{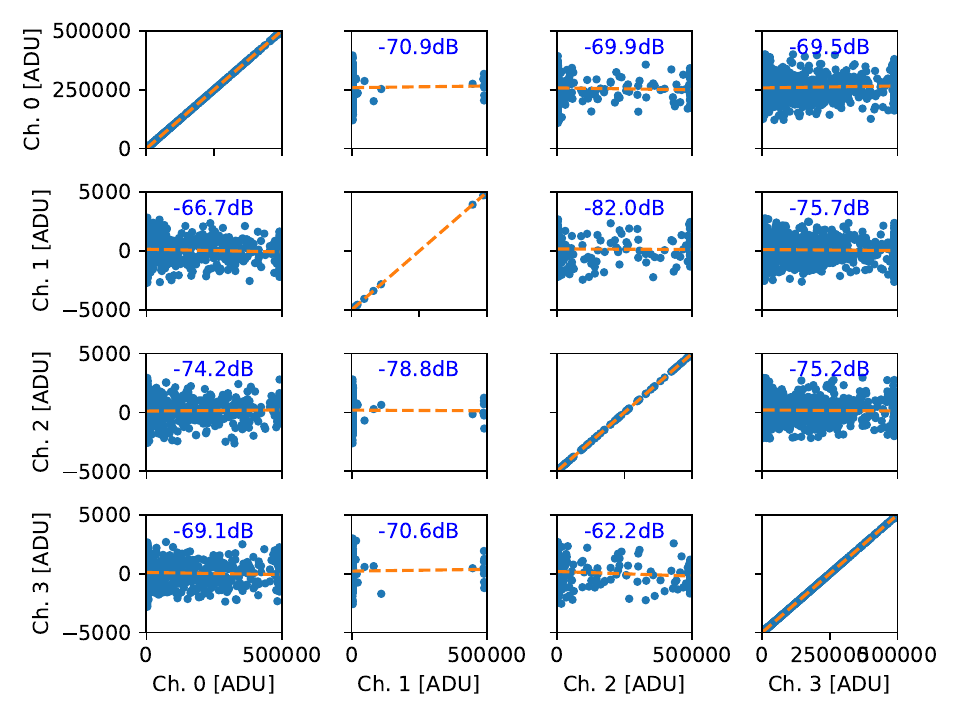}
        \label{fig:crosstalkMIDNA2_correlation}
    }
    
    \caption{Crosstalk evaluation of the present work. Although there is a large signal due to a particle interaction on channel 2, no effect can be seen on the neighboring channels.}
    \label{fig:crosstalkMIDNA2}
\end{figure}

On an equivalent measurement performed with the present chip and shown in Fig. \ref{fig:crosstalkMIDNA2}, it is not possible to visually distinguish any crosstalk from noise. In Fig. \ref{fig:crosstalkMIDNA2_images} and Fig. \ref{fig:crosstalkMIDNA2_pixelvalues}, the aggressor is on channel 2 with a signal amplitude similar to the previous case, however there is no evident \emph{ghost} on the other channels. 

%Assuming that there is a crosstalk and that it is masked by the readout noise, its magnitude is less than \SI{-52}{\dB}. For this calculation the RMS noise (standard deviation) of the victim channel was used as crosstalk amplitude on the victim side. \todo{So we can not see crosstalk because is below the noise floor...}

The scatter plots of Fig. \ref{fig:crosstalkMIDNA2_correlation} show an almost flat trend, and the correlation coefficients are close to zero, meaning that there is no clear relation between the amplitude on one channel and the amplitude in the others. The crosstalk parameters calculated from the slope of the linear regression are below \SI{-62}{\dB} for all channel combinations.

\subsection{Integrator offset measurements}

\begin{figure}
    \centering
    \includegraphics[width=\linewidth]{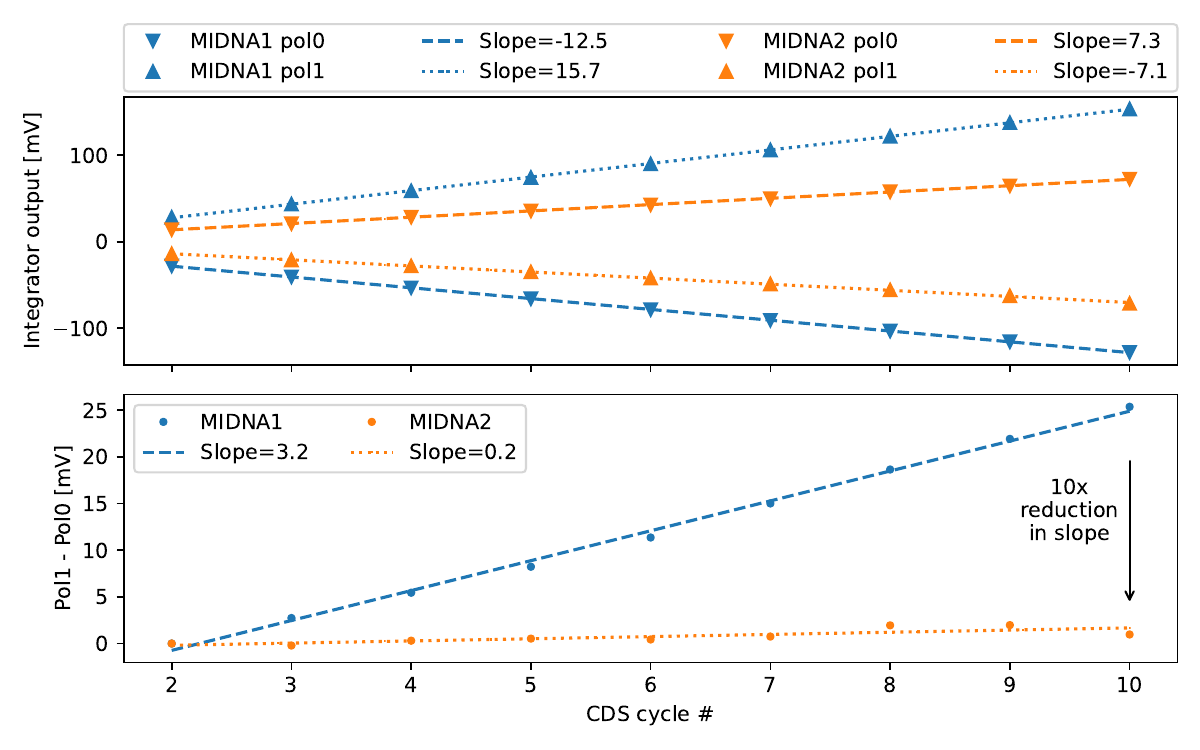}
    \caption{Measured integrator residual offset after successive CDS sequences, with the ASIC's POL input in either zero or one. The integrations were carried out with the DCR switch closed, i.e. zero signal, and with an integration time $t_{int}$ equal to \SI{13.3}{\us}.}
    \label{fig:offsetmeasurements}
\end{figure}

The integrator's offset was characterized through the following experiment: With the DCR switch always closed---S1 in Fig. \ref{fig:MIDNA1MIDNA2ChannelComparison}---several successive CDS cycles were performed, sampling the integrator output at the end of each cycle. Having the DCR switch closed means that the signal is the same for the positive and for the negative integrations, so the output is an integration of the amplifier's own offset. The experiment was carried out with POL equal to zero, equal to one, and then alternating. Both the previous and current iterations were measured in order to compare the results. Fig. \ref{fig:offsetmeasurements} shows the measured data along with a linear fit over each data set. It also plots the difference between each point with POL=1 and POL=0.

After ten successive CDS cycles with POL kept constant, the integrator output voltage of present work is approximately half of that of previous iteration, which means that the integrator's offset was reduced effectively with the changes implemented. Importantly, the symmetry between both POL states was improved a factor 10, as is shown by the slope of the POL difference plot. 

The enhanced symmetry greatly increased the number of samples that could be summed in the analog domain. For example, when using the analog pile-up sequence switching polarity between CDS cycles, if 100 samples were acquired with the previous chip at \SI{3.2}{\mV\per{}CDS}, the integrator output would add \SI{320}{\mV} of residual offset, leaving just \SI{680}{\mV} for signal. On the other hand, the chip in the present work adds only \SI{0.2}{\mV\per{}CDS}, thus the residual offset would be just \SI{20}{\mV} with almost the whole voltage range remaining for signal excursion.

\subsection{Bandgap reference measurements}\label{sec:bandgapreferencemeasurements}

\begin{figure}
    \centering
    \includegraphics[width=\linewidth]{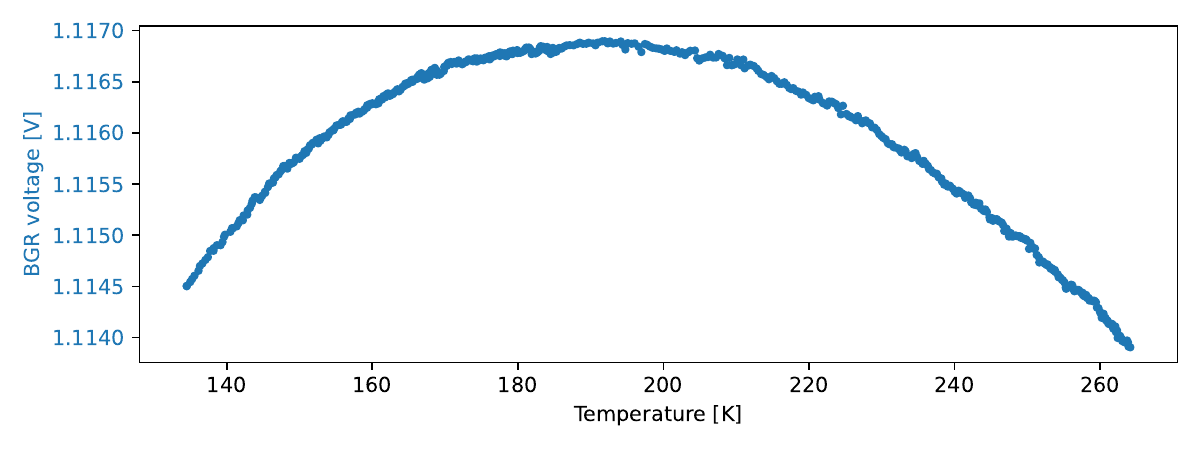}
    \caption{Measured bandgap reference voltage versus temperature. The voltage was measured at the channel output, using the integrator as a buffer, i.e. in reset mode.}
    \label{fig:BGRmeasurements}
\end{figure}

Using the same setup, the bandgap reference voltage was measured. Since a direct connection to the BGR output is not available, an indirect measurement was performed by registering the voltage at the channel output and, therefore, measuring the BGR through two buffer stages: the channel reference buffer and the integrator configured as a buffer. The channel architecture is flexible enough that it was possible to use the integrator as a buffer by keeping the reset signal activated. With the channel in this mode, 110016 samples were acquired with the LTA board, measuring the difference between the channel output and an external room-temperature \SI{1.1}{\V} reference. Fig. \ref{fig:BGRmeasurements} shows the average measured voltage, adjusted for the external reference, across a temperature range from \SI{130}{\K} to \SI{270}{\K}. The variation of approximately \SI{3}{\mV} is well within the 10 \% target.

\section{Summary and conclusions}

The present version of the MIDNA Skipper-CCD readout ASIC was presented in this work. It comprises four signal processing channels that implement a CDS technique based on dual-slope integrators. It can operate at cryogenic temperatures in close proximity to the Skipper-CCD, thus reducing the delays and noise coupling related to long wires. The channel operation is controlled by only five inputs, so the readout sequence can come from the main system controller. The inputs allow a variety of readout modes: buffer, single-sample CDS, and analog summation. In this work, the analog summation mode was optimized further to reduce the number of analog to digital conversions in large experiments like OSCURA.

Understanding the limitations identified in the previous iteration, several design changes are presented in this work. Keeping the same channel structure and control interface, the design of the ASIC improves the channel crosstalk by adding a reference voltage buffer per channel. With this addition, the channel to channel crosstalk was reduced from \SI{-38}{\dB} to \SI{-62}{\dB} according to the worst cases shown in section \ref{sec:crosstalkmeasurements}. Additionally, the integrator offset was also reduced by adding complementary switches, increasing the total capacitance of the feedback capacitor and improving the layout. In the new design the integrator offset was halved and, more importantly, it is now more symmetric when switching polarity, therefore reducing the CDS residue by ten times compared to the first iteration. Finally, the addition of an integrated bandgap reference eliminates other chips close to the sensors, which can compromise the radiopurity requirements of very sensitive physics experiments. 

The measurements presented demonstrate that in a setup with a Skipper-CCD and the presented ASIC at \SI{140}{\K} it is possible to lower the readout noise down to the single electron resolution by summing samples in the analog domain with the analog pile-up technique. The lowest readout noise shown in this work is \SI{0.11}{e^{-}_{rms}} when averaging 1200 samples.

\section*{Acknowledgements}
This work was produced by FermiForward Discovery Group, LLC under Contract No. 89243024CSC000002 with the U.S. Department of Energy, Office of Science, Office of High Energy Physics. Publisher acknowledges the U.S. Government license to provide public access under the DOE Public Access Plan.

% can use a bibliography generated by BibTeX as a .bbl file
% BibTeX documentation can be easily obtained at:
% http://www.ctan.org/tex-archive/biblio/bibtex/contrib/doc/
% The IEEEtran BibTeX style support page is at:
% http://www.michaelshell.org/tex/ieeetran/bibtex/
\bibliographystyle{IEEEtran}
% argument is your BibTeX string definitions and bibliography database(s)
\bibliography{IEEEabrv,biblio}

\end{document}